\documentclass[runningheads]{llncs}
\usepackage{graphicx}
\usepackage{amsmath}
\usepackage{hyperref}
\usepackage{booktabs}
\usepackage{subcaption}
\usepackage{adjustbox}
\hypersetup{
    colorlinks=true,   
    citecolor=black,
    urlcolor=blue,
}
\begin{document}
\title{Revealing the Hidden Impact of Top-N Metrics on Optimization in Recommender Systems}
\titlerunning{The Hidden Impact of Top-N Metrics on Optimization}
\author{
    Lukas Wegmeth\orcidID{0000-0001-8848-9434} \and
    Tobias Vente\orcidID{0009-0003-8881-2379} \and
    Lennart Purucker\orcidID{0009-0001-1181-0549}
} 
\authorrunning{L. Wegmeth, T. Vente, and L. Purucker}
\institute{
    Intelligent Systems Group, University of Siegen, Germany\\\email{\{lukas.wegmeth,tobias.vente\}@uni-siegen.de,purucker@cs.uni-freiburg.de}
}
\maketitle
\begin{abstract}
The hyperparameters of recommender systems for top-n predictions are typically optimized to enhance the predictive performance of algorithms.
Thereby, the optimization algorithm, e.g., grid search or random search, searches for the best hyperparameter configuration according to an optimization-target metric, like  \emph{nDCG} or \emph{Precision}. 
In contrast, the optimized algorithm, e.g., \emph{Alternating Least Squares Matrix Factorization} or \emph{Bayesian Personalized Ranking}, internally optimizes a different loss function during training, like \emph{squared error} or \emph{cross-entropy}. 
To tackle this discrepancy, recent work focused on generating loss functions better suited for recommender systems.
Yet, when evaluating an algorithm using a top-n metric during optimization, another discrepancy between the optimization-target metric and the training loss has so far been ignored.
During optimization, the top-n items are selected for computing a top-n metric; ignoring that the top-n items are selected from the recommendations of a model trained with an entirely different loss function. 
Item recommendations suitable for optimization-target metrics could be outside the top-n recommended items; hiddenly impacting the optimization performance.
Therefore, we were motivated to analyze whether the top-n items are optimal for optimization-target top-n metrics. 
In pursuit of an answer, we exhaustively evaluate the predictive performance of 250 \emph{selection strategies} besides selecting the top-n.
We extensively evaluate each \emph{selection strategy} over twelve implicit feedback and eight explicit feedback data sets with eleven recommender systems algorithms.
Our results show that there exist \emph{selection strategies} other than top-n that increase predictive performance for various algorithms and recommendation domains.
However, the performance of the top $\sim43\%$ of \emph{selection strategies} is not significantly different.
We discuss the impact of our findings on optimization and re-ranking in recommender systems and feasible solutions. 
The implementation of our study is publicly available.

\end{abstract}
\section{Introduction}
Top-n recommendations, i.e., recommending ranked item lists, are probably the most common task for recommender systems nowadays.
To tackle this task, recommender systems developers often apply machine learning algorithms, e.g., nearest neighbor or matrix factorization approaches \cite{portugal2018use}.
The models produced by such algorithms are then used to predict personalized ranked lists. 
These recommendations are then commonly evaluated in terms of predictive performance with optimization-target metrics like the \emph{nDCG} or \emph{Precision} \cite{10.1145/3556536,10.1145/963770.963772,zhang2016recommender,HERNANDEZDELOLMO2008790}.
This performance is influenced by the data set, the algorithm, and its hyperparameters.
Hyperparameter optimization techniques like grid search, random search, or Bayesian optimization are commonly applied to improve recommendation performance by determining the best hyperparameter values for an algorithm \cite{ekstrand2020lenskit,recbole[1.0],recbole[2.0],recbole[1.1.1],10.1145/3523227.3551472,DBLP:conf/sigir/AnelliBFMMPDN21,sun2022toward,rendle2021revisiting,10.1145/3359555.3359557}. 

No matter which optimization technique is applied, it is vital to correctly approximate the predictive accuracy of a set of hyperparameters, as this influences future decisions during the optimization or when deploying the model in production.
However, the predictive performance according to optimization-target metrics is not necessarily optimized during the training of recommender systems.
The research community knows this discrepancy between training loss metrics, e.g., \emph{squared error}, and optimization-target metrics, like \emph{nDCG}.
Their problem is finding a training loss metric that accurately represents the optimization-target metric such that the trained weights are optimized correctly.
To this end, recent work focused on empirically proven loss functions or learning better loss functions \cite{10.1145/2507157.2507210,rendle2012bpr,10.1145/3477495.3531941}.

In theory, the aforementioned discrepancy also exists in evaluation metrics, e.g., \emph{nDCG@10}.
By definition, \emph{nDCG@10} is the result of calculating the \emph{nDCG} of the top 10 predicted items, ignoring all items ranked below the top 10.
Metrics like \emph{nDCG@10} strongly differ from training loss functions that evaluate model performance without the concept of a ranking or relevance threshold.
Consequently, we hypothesize that, due to the aforementioned discrepancy between the metrics, evaluating only the top-n selection is insufficient to optimize for the highest predictive accuracy of the recommender system.
In other words, evaluating only the top-n predicted items but ignoring all others might not always yield the highest possible predictive accuracy of a trained model.

Therefore, we aim to determine whether there are cases where selecting items other than the top-n results in higher predictive accuracy. 
If there were such cases, it would imply that there is a hidden impact of top-n metrics on optimization performance due to the assumption that evaluating the top-n recommended items results in the highest predictive accuracy w.r.t. the evaluated hyperparameters.
The aforementioned implication motivated us to conduct the exploratory study that we present in this paper.
In detail, our study aims to answer the following questions:
\phantomsection
\label{rq}
\begin{enumerate}
    \item[\textbf{RQ1}] Does the selection of items other than the top-n during the evaluation of recommender systems yield improved predictive accuracy for specific algorithms, domains, or data sets?
    \item[\textbf{RQ2}] If there are cases where selecting items other than the top-n improves predictive accuracy, is there a \emph{significant} impact of top-n metrics on optimization?
\end{enumerate}

To answer the research questions, we conduct a study of the performance of nine recommendation algorithms on twelve implicit feedback and eight explicit feedback data sets.
\textbf{Our contribution} is the first large-scale study and discussion of the, so far, hidden impact of top-n metrics on optimization in recommender systems.
Our results prove that there are cases where selecting items other than the top-n yields increased performance, answering \textbf{RQ1}.
However, we indicate that the impact is likely insignificant, answering \textbf{RQ2}.
Therefore, we reveal the hidden impact of top-n metrics on optimization in recommender systems and provide evidence that researchers do not have to worry about it being a confounding factor in the evaluation of recommender systems using traditional collaborative filtering algorithms.
Our long-term goal is to clear doubts about hidden problems in evaluating recommender systems and to raise awareness in the community. 

Our implementation is publicly available on our GitHub repository\footnote{\url{https://code.isg.beel.org/scoring-optimizer}} and contains documentation for the reproducibility of our experiments.
\section{Related Work}
To the best of our knowledge, there exists no analysis of the quality of the top-n selection of ranked lists.
However, our work is related to and motivated by recent work on hyperparameter optimization, training loss metrics, and re-ranking.

The optimization of hyperparameters requires some score that approximates the highest predictive accuracy of a model.
If the score fails to do that, the optimization strategy runs into the risk of either optimizing for undesired criteria, e.g. in Bayesian optimization, or the results can not be interpreted correctly, e.g. in grid search and random search. 
Recommender systems research that reports evaluation scores usually obtains these scores on optimized hyperparameters \cite{10.1145/3523227.3546752,10.1145/3523227.3546788,10.1145/3383313.3412236}.
Additionally, there are efforts to transfer automated machine learning techniques to recommender systems \cite{10.1145/3604915.3610656,anand2020auto,10.1145/3579355,10.1145/3397271.3401436,joglekar2019neural,chen2023comprehensive}, and one of the core problems of automated machine learning is automated hyperparameter optimization \cite{NIPS2015_11d0e628}.

Related to optimization, the discrepancy between training loss and optimization-target metrics has been explored by the recommender systems community \cite{10.1145/3269206.3271784,10.1145/3341981.3344221}.
The problem can be tackled either by engineering loss metrics that fit well to the desired accuracy metric \cite{10.1145/2507157.2507210,rendle2012bpr} or by learning the loss metric automatically \cite{10.1145/3477495.3531941}.
In contrast, we thoroughly analyze the validity of selecting only the top-n elements for evaluation.

To achieve secondary optimization goals that are different from increasing predictive accuracy, e.g., removing popularity bias, re-ranking techniques are commonly applied \cite{10.1145/3397271.3401431,abdollahpouri2019managing,liu2022neural,pei2019personalized,jannach2017price}.
Re-ranking techniques assume that the underlying ranked list is predicted by an implicit feedback recommender system optimized for predictive accuracy.
However, our analysis focuses on re-ranking predicted items during the optimization of a model to better approximate the predictive accuracy of a set of hyperparameters.
Our approach is more similar to works that analyze strategies that randomly sample recommendations from the top predicted items \cite{DBLP:conf/ecir/LangerB17}, or define a relevance cutoff for recommended items \cite{DBLP:conf/ecir/BeelD17}.

The Probability Ranking Principle (PRP) \cite{ROBERTSON1977} assumes that an optimal recommender algorithm ranks items in order of probability of relevance to the user \cite{10.1145/3437963.3441662}.
In contrast, our work investigates whether a global ranking order of items is optimal on average for all users.
Hence, if the best possible global ranking order of an algorithm's predictions contains exactly the top-n predicted items, then the PRP holds on average for all users.
However, our work aims to research the impact of top-n metrics on optimization.
\section{Method}
At the core of this paper is the evaluation of different \emph{selection strategies}.
A recommender system predicts a ranked list of items $P$ for each user.
We define a \emph{selection strategy} as an approach to select any $n$ items from $P$, with $n \leq |P|$, for the evaluation of the recommender system's accuracy with a threshold-based metric like  \emph{nDCG@n}.
Hence, the number of possible \emph{selection strategies} is $\binom{|P|}{n}$.
In the traditional evaluation of recommender systems, e.g. with \emph{nDCG@10} (i.e., $n=10$), the commonly used \emph{selection strategy} is choosing the top 10 items from $P$.
We call this \emph{selection strategy} the \emph{top-n selection strategy}.
To illustrate an alternative to the \emph{top-n selection strategy}, assume we randomly select 10 items from $P$ and use this \emph{non-top-n selection strategy} to compute the \emph{nDCG@10}.
Such an approach was, for example, proposed in research on the effectiveness of randomly sampling ranked items \cite{DBLP:conf/ecir/LangerB17}. 

Our work focuses on finding a \emph{non-top-n selection strategy} with a higher \emph{nDCG@n} than the \emph{top-n selection strategy}.
Additionally, we define the subset $K$, with $K\subseteq P$, as the subset that contains the top-$|K|$ elements of $P$, which we need later.

The experiments presented in this paper were executed in a highly parallel manner on a cluster where each node has 256 GB RAM and a total of 64 cores from two AMD EPYC 7452 CPUs.

\subsection{Selection Strategy Design Decisions}
\label{selection_strategy_decision}
For our exploratory study, we set $|K|=10$ and $n=5$ to investigate the resulting 252 distinct \emph{selection strategies}.
In this paragraph, we provide our motivation for this choice.
For context, the \emph{top-n selection strategy} in this setting is choosing the top 5 predicted items from $K$.
To answer \textbf{RQ1}, technically, we only need to show that there exists at least one \emph{non-top-n selection strategy} with higher performance than the \emph{top-n selection strategy} for any combination of $|K|$ and $n$.
However, to sufficiently answer \textbf{RQ2}, we want to acquire more exhaustive results than for, e.g. $|K|=2$ and $n=1$, or $|K|=11$ and $n=10$.
Further, evaluating the lowest-ranked items does not provide additional information, e.g., when $|K|$ is close to or equal to $|P|$.
Moreover, if \emph{selection strategies} that choose the lowest-ranked items perform well, the evaluated algorithm likely fails to learn correctly from the data.
Additionally, the number of possible \emph{selection strategies} increases exponentially with $|K|$ if it significantly differs to $n$.
In the complete study, including data preprocessing, fitting, and predicting, the parameters $|K|$ and $n$ have the most considerable influence on computational resource requirements for our exploratory study.
With $|K|=10$ and $n=5$, analyzing one data set requires about one CPU year on average with our evaluation setup.
Furthermore, $n=5$ is commonly chosen for recommender systems evaluations.
Considering all the above points, we selected $|K|=10$ and $n=5$ for this study.

\subsection{Data Sets and Algorithms}
We analyzed twelve implicit and eight explicit feedback data sets from six different domains with nine recommendation algorithms plus two baselines from two recommender systems libraries.
Four of the recommendation domains, \emph{shopping}, \emph{music}, \emph{movies}, and \emph{articles}, are represented by three or more data sets each.
The remaining data sets are from the domains \emph{social} and \emph{locations}.
The implicit feedback data sets are: \emph{Adressa One Week} \cite{10.1145/3106426.3109436}, \emph{Citeulike-a} \cite{DBLP:conf/ijcai/WangCL13}, \emph{Cosmetics-Shop}\footnote{\url{https://rees46.com/}}, \emph{Globo} \cite{de_Souza_Pereira_Moreira_2018,Moreira_2019}, \emph{Gowalla} \cite{10.1145/2020408.2020579}, \emph{Hetrec-Lastfm} \cite{Cantador:RecSys2011}, \emph{Nowplaying-rs} \cite{smc18}, \emph{Retailrocket}\footnote{\url{https://www.kaggle.com/datasets/retailrocket/ecommerce-dataset}}, \emph{Sketchfab}\footnote{\url{https://github.com/EthanRosenthal/rec-a-sketch}}, \emph{Spotify-Playlists} \cite{7395827}, \emph{Yelp}\footnote{\url{https://www.yelp.com/dataset}}, and\emph{Yoochoose}\footnote{\url{https://www.kaggle.com/datasets/chadgostopp/recsys-challenge-2015}}.
The explicit feedback data sets are: \emph{Amazon CDs\&Vinyl} \cite{ni-etal-2019-justifying}, \emph{Amazon Musical Instruments} \cite{ni-etal-2019-justifying}, \emph{Amazon Video Games} \cite{ni-etal-2019-justifying}, \emph{CiaoDVD}\footnote{\url{https://guoguibing.github.io/librec/datasets.html}}, \emph{Jester3} \cite{goldberg2001eigentaste}, \emph{MovieLens-1M} \cite{10.1145/2827872}, \emph{MovieLens-100k} \cite{10.1145/2827872}, and \emph{MovieTweetings} \cite{4284240}.
For these explicit feedback data sets, we treat a rating that is $>60\%$ of the maximum rating as an interaction according to standard practice \cite{10.1145/3459637.3482056,10.1145/3523227.3546756,10.1145/3178876.3186150}.
Furthermore, we prune all data sets such that all users and items have at least five interactions, commonly called five-core pruning \cite{10.1145/3357384.3357895,10.1145/3460231.3474275,10.1145/3523227.3546770}.
We do this to reduce the impact of cold start cases since the used algorithms can not predict cold start scenarios.
Table \ref{data_statistics} contains the data set statistics for the preprocessed data sets.
We used the libraries \emph{Implicit} \cite{frederickson2018fast} and \emph{LensKit} \cite{ekstrand2020lenskit} for their implementation of the recommender algorithms.
We used all algorithms from these libraries that natively support implicit feedback for a total of nine algorithms plus two baselines.
The algorithms from \emph{Implicit} are \emph{Alternating Least Squares}, \emph{Logistic Matrix Factorization}, \emph{Bayesian Personalized Ranking}, and \emph{Item-Item Nearest Neighbors} with distance metrics \emph{Cosine Similarity}, \emph{TF-IDF}, and \emph{BM25}. 
The algorithms from \emph{LensKit} are \emph{Implicit Matrix Factorization}, \emph{User-User Nearest Neighbors}, \emph{Item-Item Nearest Neighbors}, \emph{Most Popular}, and \emph{Random}.

\subsection{Experimental Pipeline}
We perform five-fold cross-validation by randomly splitting the interactions of each user into three separate sets with fixed sizes: training (60\%), validation (20\%), and test (20\%).
All recommenders were optimized on the validation set with random search for two hours.
The range of hyperparameter values in the configuration space is set to be around the default values given by the library.
Finally, we exhaustively evaluate all 252 \emph{selection strategies} for both the validation and test data.
We focus on and present the results from evaluating \emph{selection strategies} on the test data.
However, we also obtain the results from evaluating \emph{selection strategies} on the validation data to analyze the generalization capabilities of \emph{selection strategies}.

\begin{table}[htp]
    \centering
    \caption{Data set statistics after five-core pruning. Split between the implicit (first part) and explicit (second part) feedback data sets.}  
    \resizebox{\textwidth}{!}{%
   \begin{tabular}{l|l|l|l|l|l|l|l}
    \toprule
Name & \#Interactions &     \#Users &   \#Items & Avg.\#Int./User & Avg.\#Int./Item & Sparsity &     Domain \\
\midrule
Adressa One Week           &     2,020,328 &    146,635 &    2,441 &          13.78 &         827.66 &   99.44\% &   Articles \\
Citeulike-a                &       200,180 &      5,536 &   15,429 &          36.16 &          12.97 &   99.77\% &   Articles \\
Cosmetics-Shop             &     4,949,482 &    230,248 &   44,009 &           21.5 &         112.47 &   99.95\% &   Shopping \\
Globo                      &     2,482,163 &    157,926 &   11,832 &          15.72 &         209.78 &   99.87\% &   Articles \\
Gowalla                    &     2,018,421 &     64,115 &  164,532 &          31.48 &          12.27 &   99.98\% &  Locations \\
Hetrec-Lastfm              &        71,355 &      1,859 &    2,823 &          38.38 &          25.28 &   98.64\% &      Music \\
Nowplaying-rs              &     2,447,318 &     64,392 &   95,277 &          38.01 &          25.69 &   99.96\% &      Music \\
Retailrocket               &       240,938 &     22,178 &   17,803 &          10.86 &          13.53 &   99.94\% &   Shopping \\
Sketchfab                  &       547,477 &     25,655 &   15,274 &          21.34 &          35.84 &   99.86\% &     Social \\
Spotify-Playlists          &     8,718,742 &     15,146 &  337,256 &         575.65 &          25.85 &   99.83\% &      Music \\
Yelp                       &     3,999,684 &    268,658 &  109,340 &          14.89 &          36.58 &   99.99\% &  Locations \\
Yoochoose                  &    10,195,058 &  1,283,296 &   27,995 &           7.94 &         364.17 &   99.97\% &   Shopping \\
    \midrule
   Amazon CDs\&Vinyl           &     1,241,336 &     98,228 &   66,979 &          12.64 &          18.53 &   99.98\% &   Shopping \\
Amazon Musical Instruments &       176,631 &     21,420 &    8,642 &           8.25 &          20.44 &    99.9\% &   Shopping \\
Amazon Video Games         &       362,996 &     40,793 &   14,497 &            8.9 &          25.04 &   99.94\% &   Shopping \\
CiaoDVD                    &        23,467 &      1,582 &    1,788 &          14.83 &          13.12 &   99.17\% &     Movies \\
Jester3                    &       777,118 &     34,884 &      140 &          22.28 &       5,550.84 &   84.09\% &     Social \\
MovieLens-1M               &       835,789 &      6,038 &    3,307 &         138.42 &         252.73 &   95.81\% &     Movies \\
MovieLens-100k             &        81,697 &        943 &    1,203 &          86.64 &          67.91 &    92.8\% &     Movies \\
MovieTweetings             &       563,309 &     20,643 &    8,810 &          27.29 &          63.94 &   99.69\% &     Movies \\
\bottomrule
    \end{tabular}
    }
    \label{data_statistics}
\end{table}
\section{Results}
Our results are presented in the following order.
We first show \textbf{aggregated} evaluation results of all twelve implicit and eight explicit data sets with all eleven algorithms.
Next, we present a closer look using a \textbf{domain-specific} subset of the results. 
Zooming further into this subset, we provide a more detailed, \textbf{data set-specific}, view of the behavior of different \emph{selection strategies}.
Moreover, we analyze the \textbf{generalization capabilities} of \emph{selection strategies}.
Finally, we perform \textbf{statistical tests} on the significance of \emph{selection strategies} to understand the impact of potential solutions.
All results are split between implicit and explicit feedback data sets to compare them.

Though we aggregated results as much as possible, we can only show exemplary results for some parts of the analysis in this paper due to space constraints.
The results we can not show in this paper do not provide any additional insight and do not contradict the presented results.
For example, all data presented in this paper feature results exclusively on the \emph{nDCG} metric.
However, we also performed the same evaluation procedure on the \emph{Precision} metric and found no difference compared to the results of the \emph{nDGC} metric. 
This applies to algorithms, domains, and data sets as well.
To confirm, the interested reader may refer to the complete set of results of the whole evaluation procedure, which are stored in our public repository.
Any additional material is therefore made available only for completeness and reproducibility purposes.

\begin{figure}[!ht]   
    \begin{subfigure}[b]{1\textwidth}
        \centering
        \includegraphics[width=1\linewidth]{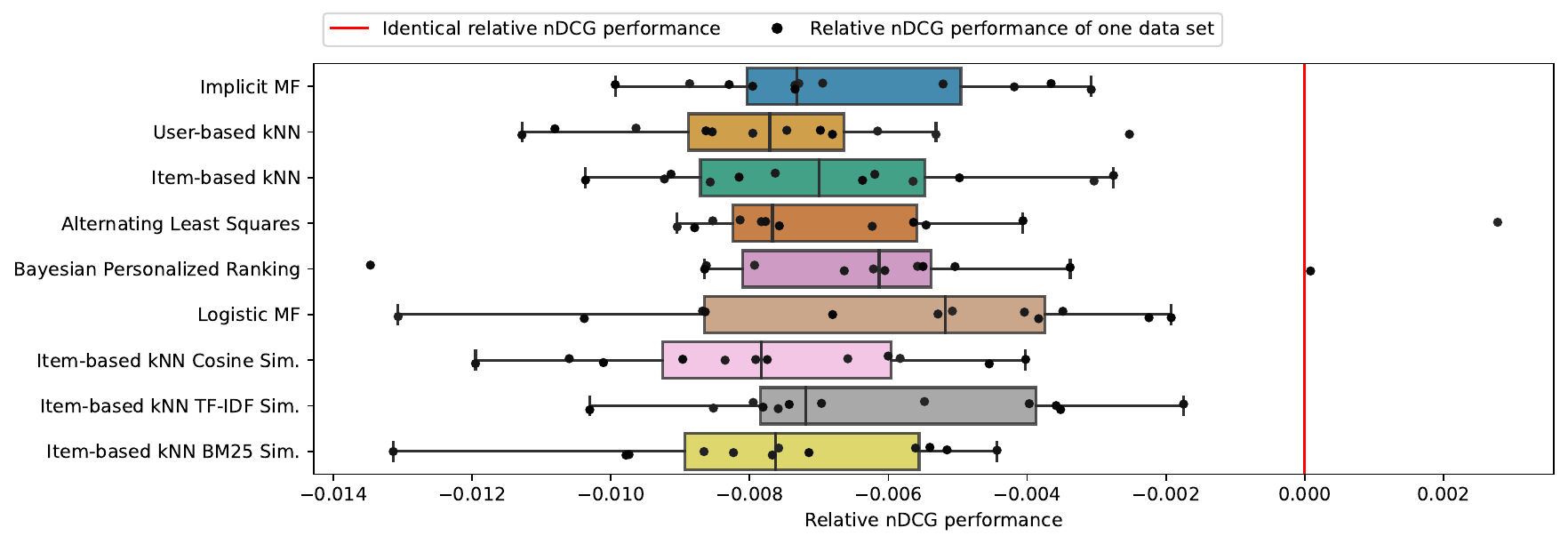}
        \caption{Non-baseline algorithms for implicit feedback.}
        \label{implicit_performance}
    \end{subfigure}\hfill
    \begin{subfigure}[b]{1\textwidth}
        \raggedleft
        \includegraphics[width=.875\linewidth]{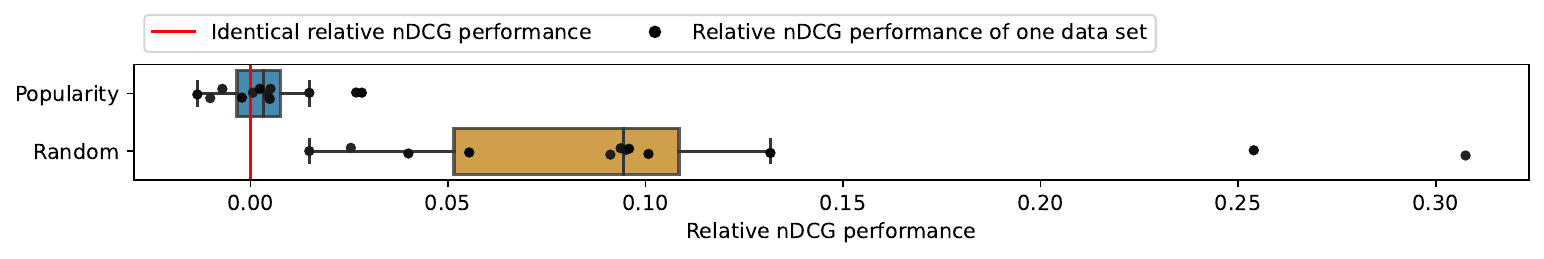}
        \caption{Baseline algorithms for implicit feedback.}
        \label{implicit_baseline}
    \end{subfigure}\hfill
    \begin{subfigure}[b]{1\textwidth}
        \centering
        \includegraphics[width=1\linewidth]{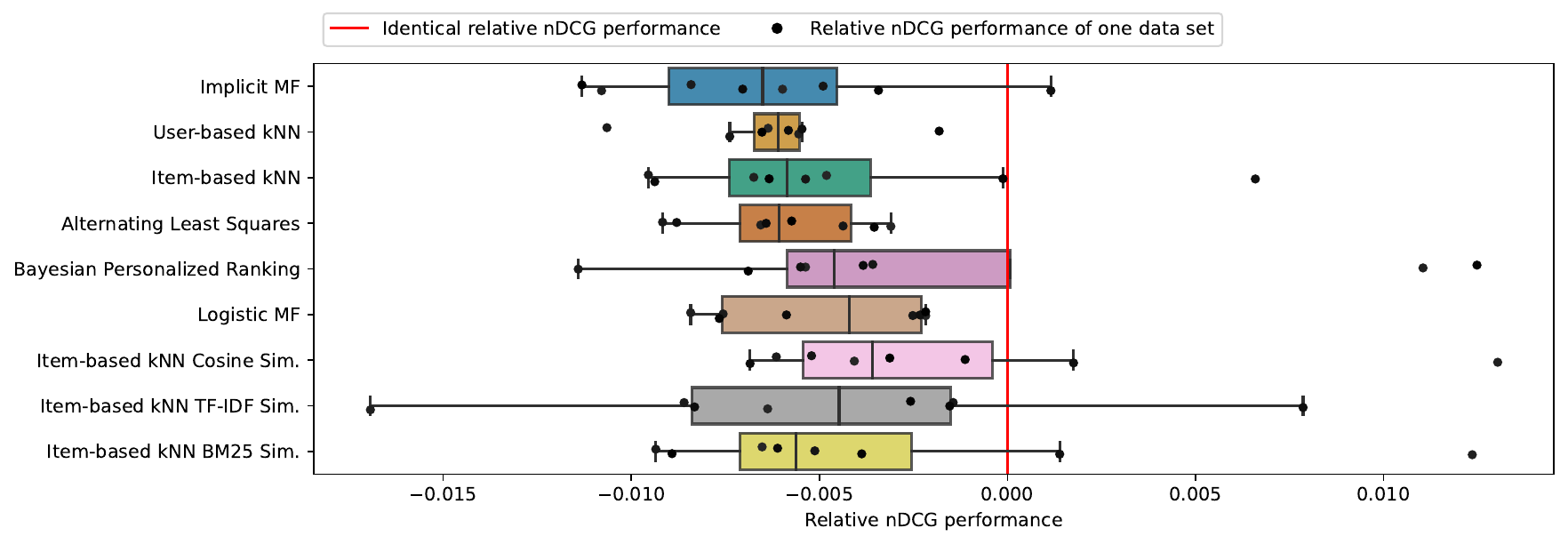}
        \caption{Non-baseline algorithms for explicit feedback.}
        \label{explicit_performance}
    \end{subfigure}\hfill
     \begin{subfigure}[b]{1\textwidth}
        \raggedleft
        \includegraphics[width=.875\linewidth]{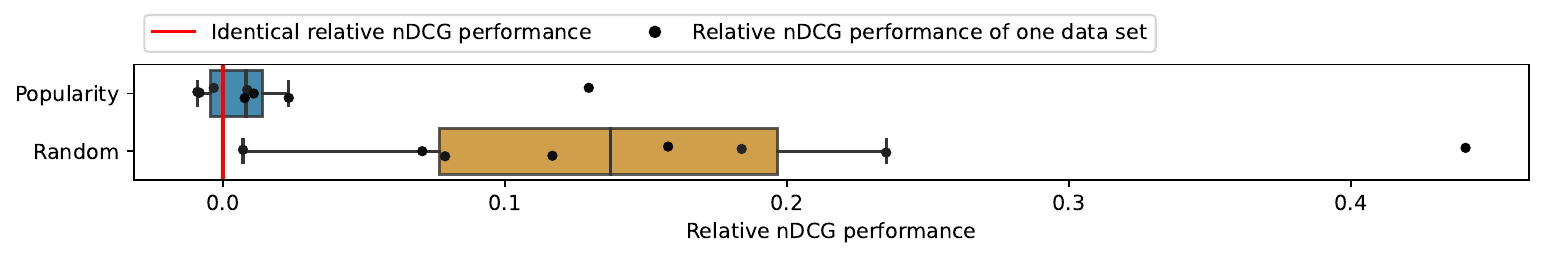}
        \caption{Baseline algorithms for explicit feedback.}
        \label{explicit_baseline}
    \end{subfigure}\hfill   
    \caption{The relative \emph{nDCG} performance of the best \emph{non-top-n selection strategy} versus the \emph{top-n selection strategy} evaluated on the test set. 
    A point to the right of the red line indicates that the best \emph{non-top-n selection strategy} is better than the \emph{top-n selection strategy}.}
    \label{aggregated_results}
\end{figure}

\paragraph{\textbf{Aggregated Results}}
Figure \ref{aggregated_results} provides an aggregated view of our results on all twelve implicit and eight explicit data sets split between implicit and explicit feedback with all eleven algorithms. 
We observe that there are some data sets and algorithms for which the best \emph{nDCG} performance is not achieved with the \emph{top-n selection strategy}.
However, the distribution of the data points also shows that the \emph{top-n selection strategy} is \emph{the best on average}.
Furthermore, with the baseline plot, we observe that the random and popularity baselines behave as expected.
When recommending random elements, there is a high probability that the \emph{top-n selection strategy} is sub-optimal due to the amount of other possible \emph{selection strategies}.
Recommending items based on popularity using the \emph{top-n selection strategy} is also sub-optimal in most cases, simply confirming that the most popular recommendations are not automatically the best.

In all cases where a \emph{non-top-n selection strategy} performs better than the \emph{top-n selection strategy}, we could search for the best \emph{non-top-n selection strategy} to increase performance.
The tested recommendation algorithms appear to be stable in terms of relative performance since the difference in performance in both directions is marginal at less than 0.4\% for implicit feedback and less than 1.5\% for explicit feedback, indicating that finding the best \emph{selection strategy} only has a marginal performance impact. 

\begin{figure}[!ht]   
    \begin{subfigure}[b]{1\textwidth}
        \centering
        \includegraphics[width=1\linewidth]{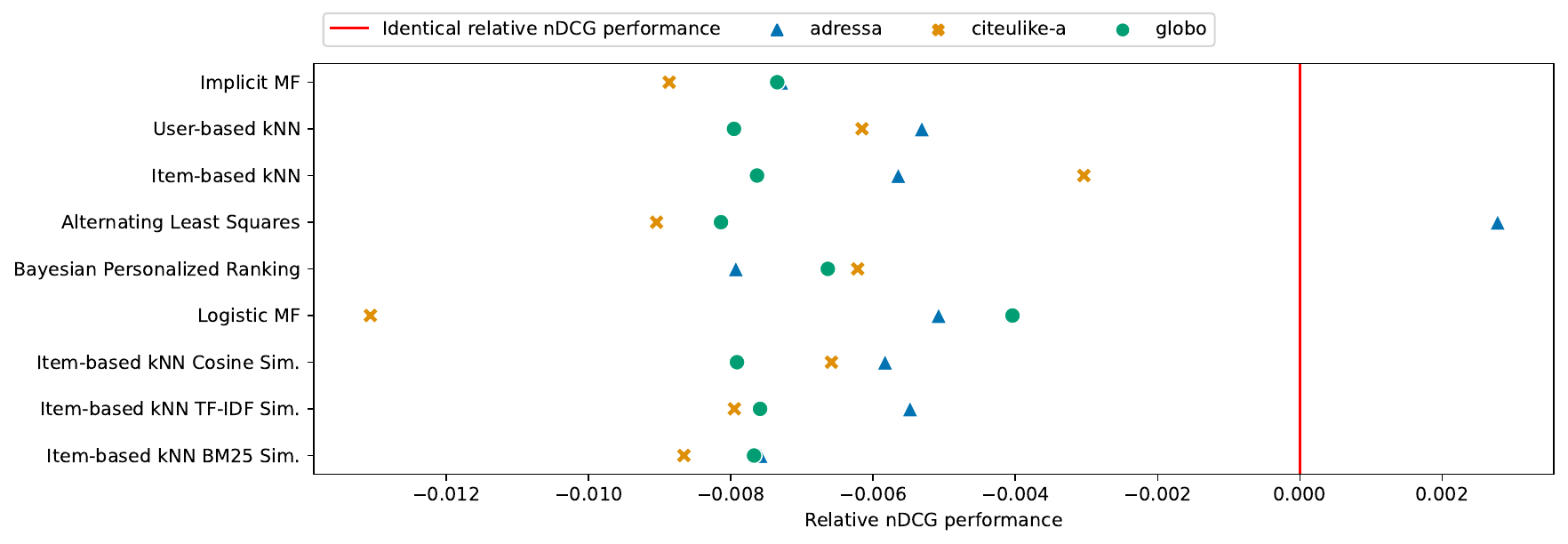}
        \caption{\emph{Articles} recommendation domain (3 data sets, implicit feedback).}
        \label{articles_domain}
    \end{subfigure}\hfill
    \begin{subfigure}[b]{1\textwidth}
       \centering
        \includegraphics[width=1\linewidth]{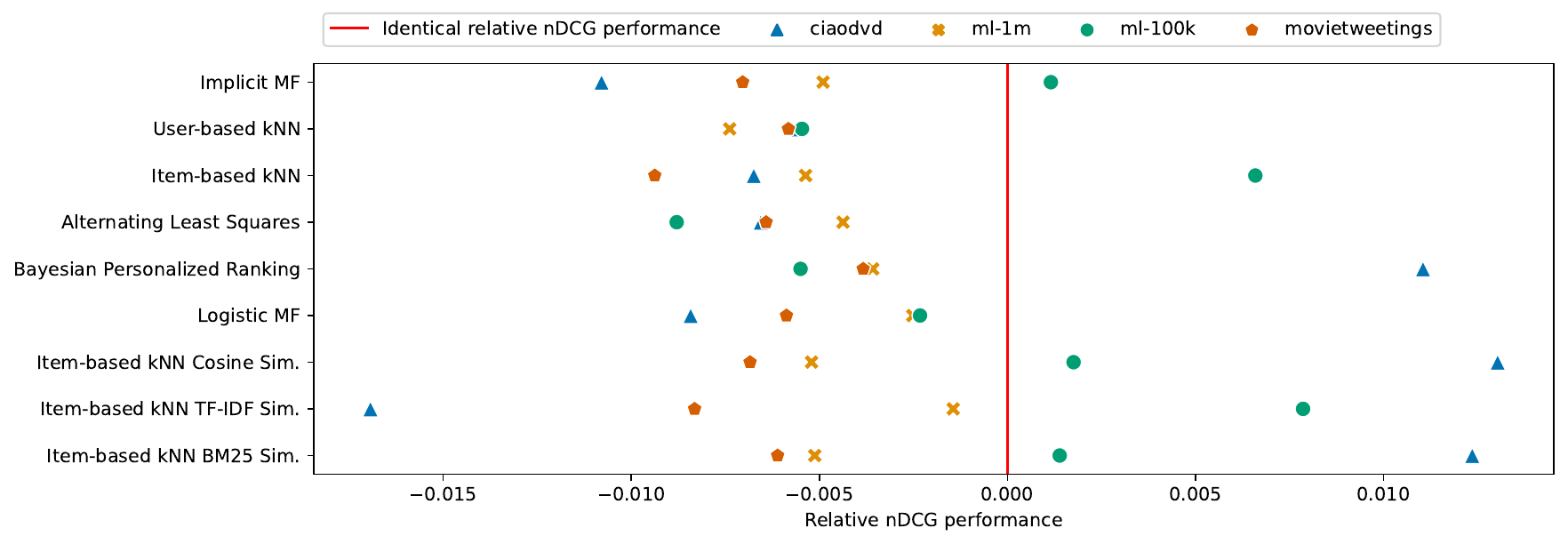}
        \caption{\emph{Movies} recommendation domain (4 data sets, explicit feedback).}
        \label{movies_domain}
    \end{subfigure}\hfill
    \caption{This Figure is a focused view on evaluating data sets that represent the \emph{articles} (\ref{articles_domain}) and \emph{movies} (\ref{movies_domain}) recommendation domains. The shown data points are a subset of Figures \ref{implicit_performance} and \ref{explicit_performance}, respectively, but with a focus on the data sets. The data sets are indicated with markers and colors to distinguish them better.}
    \label{domains}
\end{figure}

\paragraph{\textbf{Domain-Specific Results}}
To further analyze the observations, we focus on the \emph{articles} and \emph{movies} domains, which contain only implicit and explicit data sets, respectively, in Figure \ref{domains}.
Notably, there are cases with different data sets in these domains where the best \emph{non-top-n selection strategy} is better than the \emph{top-n selection strategy}.
Therefore, the aggregated and domain-specific plots indicate that the relative performance of the best \emph{selection strategies} may be specific to an algorithm or data set rather than a recommendation domain. 
Additionally, the average difference in the relative performance of data sets changes for different recommender system algorithms.

\paragraph{\textbf{Data Set-Specific Results}}
To better understand the scope of the previous observations, we have to analyze not only the best \emph{non-top-n selection strategy} but all of them. 
We, therefore, take an exemplary look at the exhaustive evaluation of \emph{selection strategies} for the \emph{Adressa One Week} data set from the \emph{articles} domain on the \emph{Alternating Least Squares} algorithm for implicit feedback.
Furthermore, we do the same for the \emph{MovieLens-100k} data set from the \emph{movies} domain on the \emph{Item-Item Nearest Neighbors} algorithm for explicit feedback.

\begin{figure}[!ht]   
    \begin{subfigure}[b]{1\textwidth}
        \centering
        \includegraphics[width=1\linewidth]{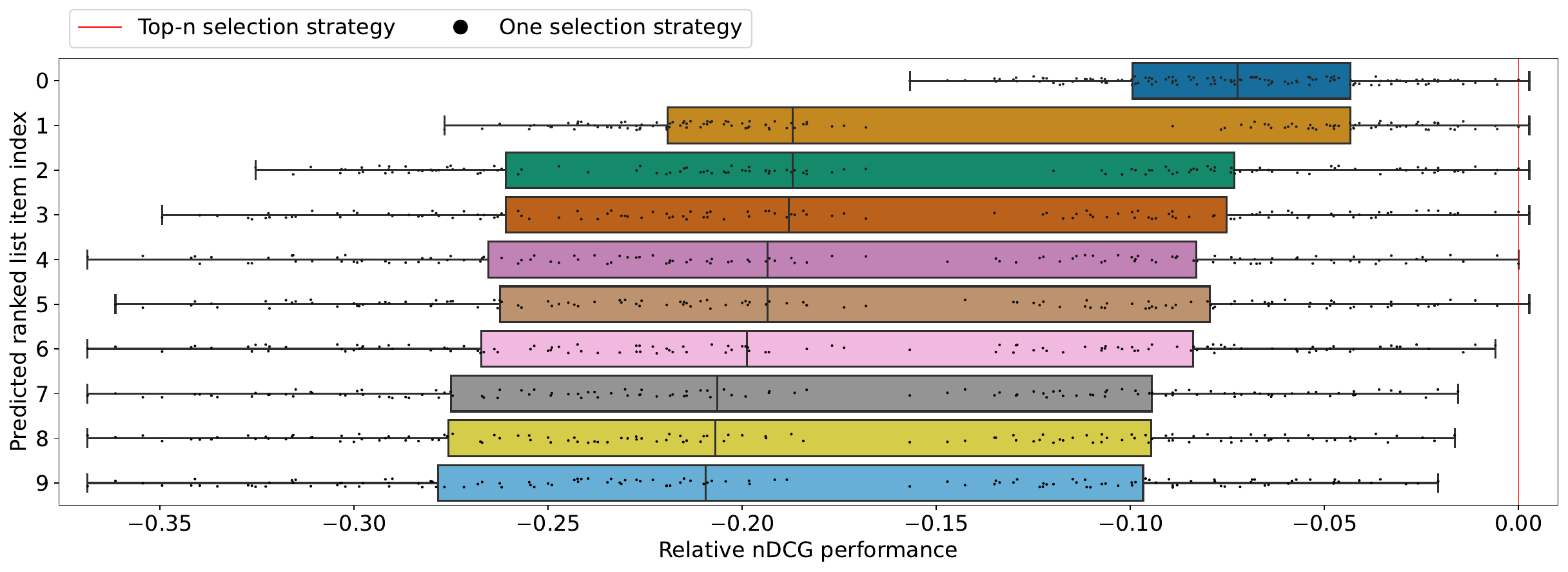}
        \caption{\emph{Adressa One Week} data set.}
        \label{adressa}
    \end{subfigure}\hfill
    \begin{subfigure}[b]{1\textwidth}
       \centering
        \includegraphics[width=1\linewidth]{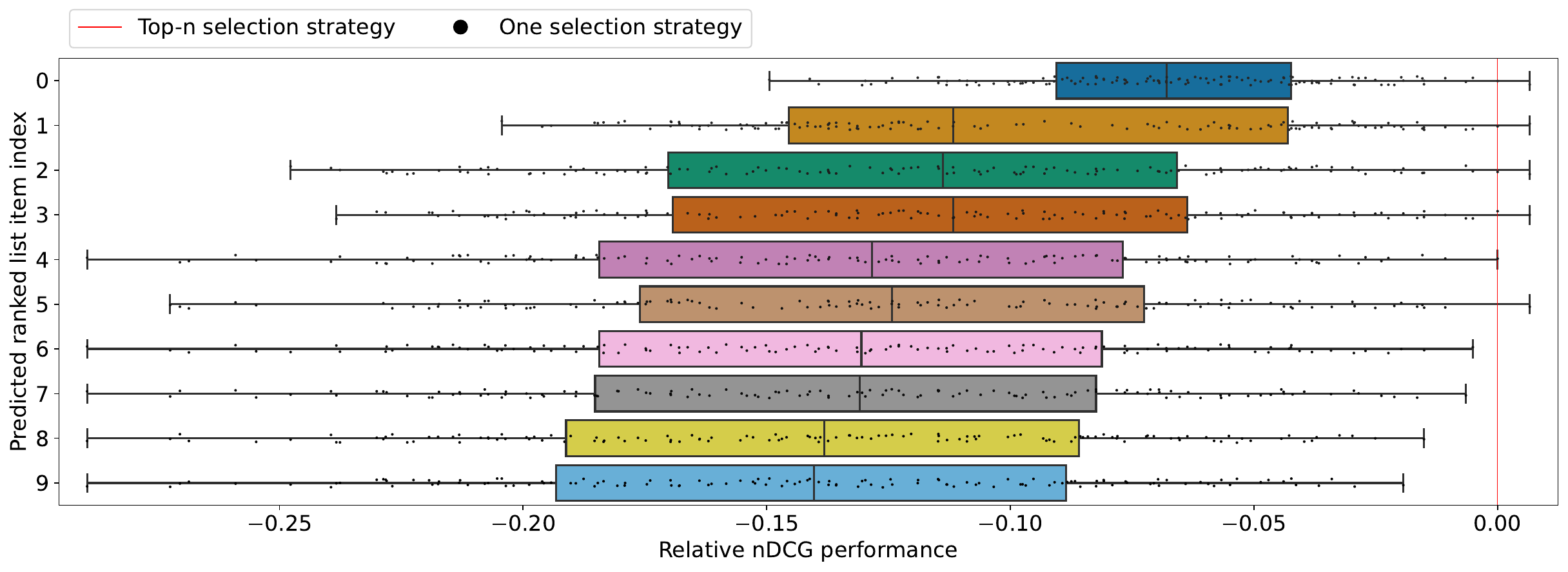}
        \caption{\emph{MovieLens-100k} data set.}
        \label{ml100k}
    \end{subfigure}\hfill
    \caption{The exhaustive evaluation of 252 \emph{selection strategies} with the \emph{nDCG} metric on the data set \emph{Adressa One Week} with the \emph{Logistic Matrix Factorization} algorithm (\ref{adressa}) and the data set \emph{MovieLens-100k} with the \emph{Alternating Least Squares} algorithm (\ref{ml100k}). Each \emph{selection strategy} picks a different combination of 5 items from the top 10 predicted items. The performance of a \emph{selection strategy} is averaged over all users in the test data set. The y-axis refers to the index of the items in the predicted ranked list, e.g., element 0 is the highest-ranked and the most relevant item. A black dot refers to a \emph{selection strategy} that contains said item, e.g., all \emph{selection strategies} represented by black dots in a row contain the item according to the index stated on the y-axis.}
    \label{selection_index}
\end{figure}

Figure \ref{selection_index} visualizes the results of the exhaustive evaluation through box plots.
If the algorithm ranks the items correctly, we expect a consistent movement toward a lower median for lower-ranked elements, but there is no such trend.
In fact, the interquartile range shows that the model only appears to be consistently performing well for the first predicted item.
However, there are items for which the plot shows increased average performance compared to a higher-ranked element.
Additionally, we already know from Figures \ref{articles_domain} and \ref{movies_domain}, specifically for these examples, that there is at least one \emph{non-top-n selection strategy} that improves over the \emph{top-n selection strategy}.
Figure \ref{selection_index} reveals the elements chosen in these strategies, indicated by the points to the right of the red vertical line, e.g., the line indicating the performance of the \emph{top-n selection strategy}.

\begin{figure}[!ht]   
    \begin{subfigure}[b]{0.45\textwidth}
        \centering
        \includegraphics[width=1\linewidth]{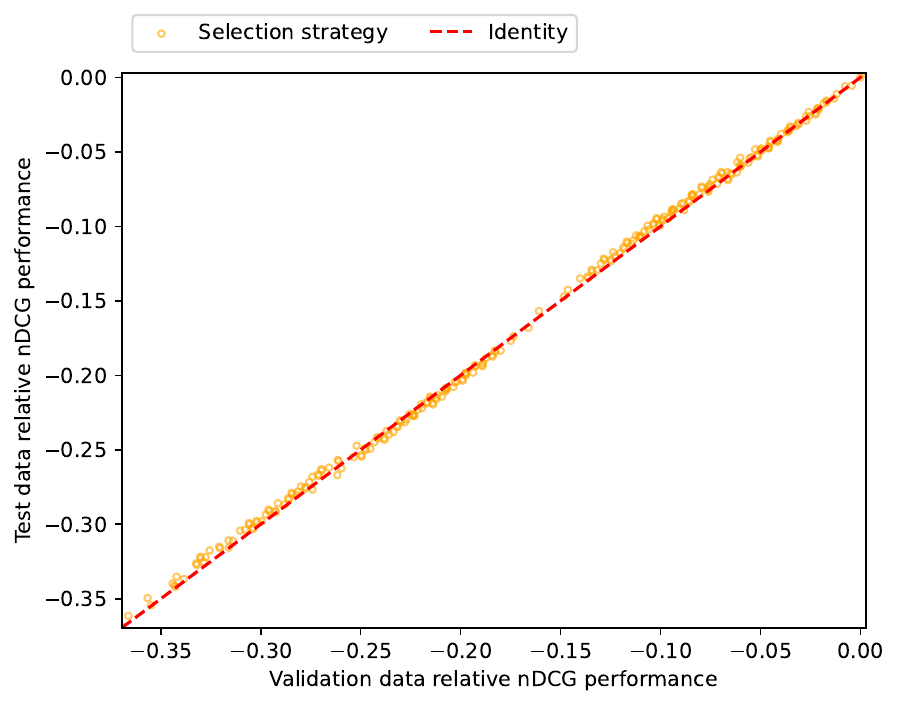}
        \caption{\emph{Adressa One Week}}
        \label{adressa_generalization}
    \end{subfigure}\hfill
     \begin{subfigure}[b]{0.45\textwidth}
        \centering
        \includegraphics[width=1\linewidth]{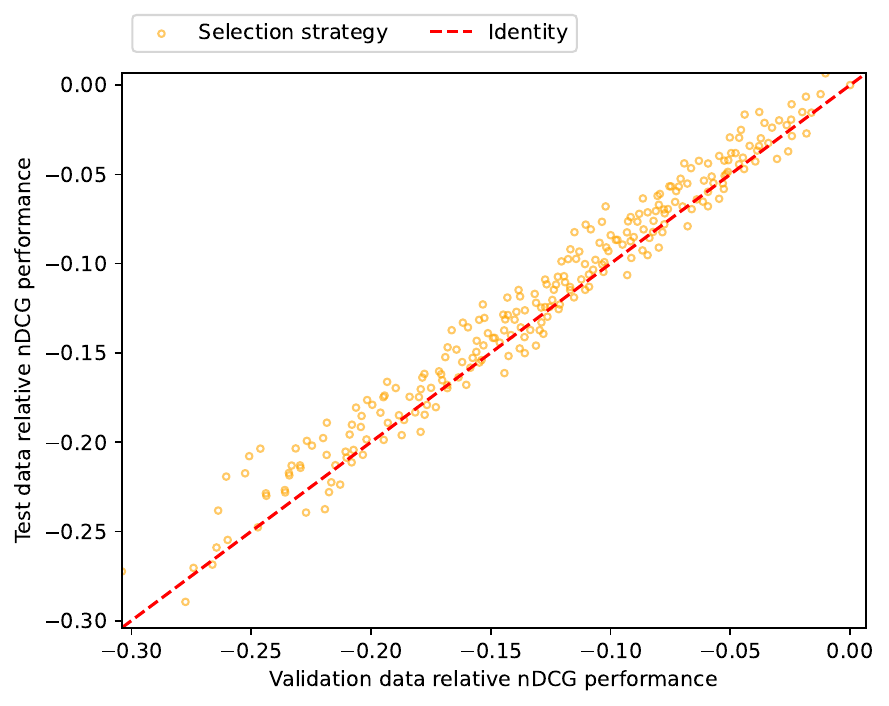}
        \caption{\emph{MovieLens-100k}}
        \label{ml100k_generalization}
    \end{subfigure}\hfill
     \caption{The relative performance of \emph{selection strategies} on the validation and test set for the data sets \emph{Adressa One Week} (\ref{adressa_generalization}) and \emph{MovieLens-100k} (\ref{ml100k_generalization}). If a point is on the identity line, the validation and test set have equal relative performance and, therefore, generalize.}
    \label{generalization_plot}
\end{figure}

\paragraph{\textbf{Generalization Capabilities}}
Since we can show that there are \emph{selection strategies} that are better than the \emph{top-n selection strategy}, we would need a way to find them reliably.
All of the results so far are evaluated on the test set, e.g., the data split that we did not know during training.
As a result, if we want to find the best \emph{selection strategy} in a real-world scenario, we would need to search for the best \emph{selection strategy} on the validation set and hope that it generalizes to the test set.
To that end, we performed the same exhaustive evaluation on the validation set to observe the generalization capability of the \emph{selection strategies}.
Continuing with the previously chosen example data set and recommendation algorithm, we show the generalization capability in Figure \ref{generalization_plot}.
With the generalization plot, we already see a trend regarding the generalization capability of \emph{selection strategies}.
To measure the generalization capability directly, we additionally calculated the Pearson correlation coefficient of the performance of \emph{selection strategies} over the validation and the test.
A higher Pearson correlation coefficient indicates higher generalization capability.
The Pearson correlation coefficient averaged over all data sets per algorithm is shown in Table \ref{correlation}.

\begin{table}[!ht]
    \caption{The Pearson correlation coefficients of the performance of \emph{selection strategies} of the validation set compared to the test set separated by implicit and explicit feedback sets. A Pearson correlation coefficient of one would mean that \emph{selection strategies} perfectly generalize from validation to test.}  
    \centering   
    \adjustbox{width=0.9\textwidth}{
        \begin{tabular}{l|c|c}
        Algorithm & Pearson Corr. Coeff. Implicit & Pearson Corr. Coeff. Explicit \\
        \hline
        Implicit MF                   & 0.998 & 0.988  \\
        User-based kNN                & 0.999 & 0.990  \\
        Item-based kNN                & 0.998 & 0.994  \\
        Alternating Least Squares     & 0.999 & 0.986  \\
        Bayesian Personalized Ranking & 0.995 & 0.966  \\
        Logistic MF                   & 0.995 & 0.983  \\
        Item-based kNN Cosine Sim.    & 0.999 & 0.990  \\
        Item-based kNN TF-IDF Sim.    & 0.998 & 0.992  \\
        Item-based kNN BM25 Sim.      & 0.999 & 0.992  \\
        Random                        & 0.004 & -0.064 \\
        Popularity                    & 0.994 & 0.995 
        \end{tabular}    
        \label{implicit_correlation}
    }
    \label{correlation}
\end{table}

\paragraph{\textbf{Statistical Significance}}
Finally, we tested for the statistical difference in the performance of \emph{selection strategies} over all data sets but per algorithm.
We applied the Friedman test to confirm that different \emph{selection strategies} result in statistically significant performance differences ($p<0.05$).
Then, we applied the Nemenyi post hoc test to obtain the critical difference.
The result shows that for all non-baseline algorithms, the best $\sim43\%$ \emph{selection strategies} are not significantly different in performance.
We note, however, that the theory behind the Nemenyi test may not hold up due to the high number of compared methods. 
Nevertheless, we believe it to be a worthwhile indicator to answer our research questions and beyond.
\section{Discussion}
First, we consider our \hyperref[rq]{\textrightarrow research questions}.
To answer \textbf{RQ1}, we evaluated all possible \emph{selection strategies} that choose 5 out of the top 10 predicted items.
We accept \textbf{RQ1} since we found that $\sim0.01\%$ of \emph{non-top-n selection strategies} result in higher performance than the \emph{top-n selection strategy}.
However, we can not identify specific criteria that lead to this effect, e.g. it occurred in different types of algorithms and data sets from different domains.
To answer \textbf{RQ2}, we additionally tested the statistical significance of \emph{selection strategies} based on their performance.
The tests indicate that most \emph{selection strategies} are not significantly different.
Moreover, the maximum performance gain is marginal at less than 0.4\% for implicit feedback data sets and less than 1.5\% for explicit feedback data sets, as shown in the aggregated results (Figure \ref{aggregated_results}).

\paragraph{\textbf{Searching for the best selection strategy}}
The average Pearson correlation coefficient of $>0.99$ for implicit feedback data sets and $>0.96$ for explicit feedback sets for all non-baseline algorithms shows that \emph{selection strategies} generalize their performance from validation to test.
Therefore, it is possible to reliably search and find the best \emph{selection strategy} on the validation set.
However, the search strategy we used for the evaluation, e.g., exhaustively evaluating all possible options, is not cost-efficient.  
We have also shown that the \emph{top-n selection strategy} is \emph{the best on average}.
Therefore, searching for the best \emph{selection strategy} in practice is only feasible with a highly efficient search strategy or when the potential for a minor improvement outweighs the cost of the search.
In future work, finding the best \emph{selection strategy} could be accomplished with an efficient search algorithm, e.g., a greedy search.
Alternatively, our results indicate that we do not have to observe items too far away from the top-n, e.g., we searched \emph{selection strategies} with 5 items out of a sample range of 10 items, but 8 items may have already been enough to find the optimum.
As a result, an exhaustive search on fewer combinations may be feasible.

\paragraph{\textbf{The impact on hyperparameter optimization}}
Hyperparameter optimization techniques rely on the developer to correctly approximate the predictive accuracy of a model to find the best hyperparameters.
If the \emph{top-n selection strategy} does not achieve that, due to being sub-optimal in terms of performance, we should avoid using it.
However, our statistical tests indicate that optimizing the \emph{selection strategy} likely has no practical impact since the evaluated performance with the majority of \emph{selection strategies} is not significantly different.
In turn, this means we may use most \emph{selection strategies} that choose 5 out of 10 items in practice.

\paragraph{\textbf{The impact on re-ranking}}
Re-ranking algorithms change which items are recommended, e.g. they apply a different \emph{selection strategy}.
Often, this means that predictive accuracy performance, e.g. \emph{nDCG} performance, is sacrificed to gain increased performance in secondary metrics, e.g. \emph{diversity} or \emph{fairness}.
Our results show that, when the re-ranking algorithm is confined to elements close to the top-n, the re-ranked predictions may actually not be significantly different in terms of predictive accuracy.

\paragraph{\textbf{Conclusion}}
We reveal the hidden impact of top-n metrics on optimization in recommender systems and show that it is insignificant.
As a result, there is no practical benefit in optimizing \emph{selection strategies}.
Conclusively, our exploratory study cleared any doubts on this confounding factor for the evaluation and reproducibility of traditional collaborative filtering algorithms.
\section*{Acknowledgements}
The OMNI cluster of the University of Siegen was used to compute the results presented in this paper.
\bibliographystyle{splncs04}
\bibliography{project}
\end{document}